# Focusing and Compression of Ultrashort Pulses through Scattering Media

**Ori Katz, Yaron Bromberg, Eran Small, Yaron Silberberg***

*Department of Physics of Complex Systems, Weizmann Institute of Science, Rehovot 76100, Israel*

*\*yaron.silberberg@weizmann.ac.il*

**Abstract:**

Light scattering in inhomogeneous media induces wavefront distortions which pose an inherent limitation in many optical applications. Examples range from microscopy and nanosurgery to astronomy. In recent years, ongoing efforts have made the correction of spatial distortions possible by wavefront shaping techniques. However, when ultrashort pulses are employed scattering induces temporal distortions which hinder their use in nonlinear processes such as in multiphoton microscopy and quantum control experiments. Here we show that correction of both spatial and temporal distortions can be attained by manipulating only the spatial degrees of freedom of the incident wavefront. Moreover, by optimizing a nonlinear signal the refocused pulse can be shorter than the input pulse. We demonstrate focusing of 100fs pulses through a 1mm thick brain tissue, and 1000-fold enhancement of a localized two-photon fluorescence signal. Our results open up new possibilities for optical manipulation and nonlinear imaging in scattering media.

The propagation of light in inhomogeneous media results in scattering and distortions of the propagating wavefront. Such distortions limit the effective focusing of optical intensity and degrade imaging quality through disordered or scattering media[1]. The problem of focusing light through inhomogeneous media is even more challenging when ultrashort pulses are considered, as in addition to the spatial distortions scattering also distorts the pulse shape in time[2-5]. The challenge of correcting the spatial distortions induced by scattering has been in the focus of many recent works[5-19]. Weak wavefront aberrations, such as those occurring in astronomical observations through the atmosphere, have been efficiently corrected using adaptive optics techniques[6-8]. These techniques, however, were considered inadequate for correcting distortions in highly scattering and turbid samples, which lead to diffusive light propagation and result in complex speckle patterns with no simple relation to the incident wavefront[1, 20]. Recently, in a pioneering work, Vellekoop et al. have shown that adaptive optimization of the incident wavefront can increase the focused intensity of multiply scattered light by a factor that is roughly equivalent to the number of degrees of control[9-12]. Using a spatial light modulator (SLM) with 1000 degrees of control enabled a 1000-fold enhancement in the focused brightness after a turbid medium[9]. Following Vellekoop's works, other approaches for determining the optimal corrections were demonstrated either by measurement of the optical transmission matrix[13, 14] or the complex-valued relation between spatial modes[15], or alternatively by directly recording the distorted wavefront using optical phase conjugation[16, 17].

These results, however, were only relevant for quasi-continuous light, and in spite of these remarkable achievements in the correction of spatial distortions, no work to date has addressed the simultaneous correction of the *temporal* distortions which become important when ultrashort pulses are employed[2-5]. Understanding and correcting the spatio-temporal distortions of ultrashort pulses is crucial when multiphoton processes are involved, as is the case in nonlinear microscopy and quantum coherent control experiments[5, 21-28]. Moreover, as the signal in an N-photon process is proportional to the N-th power of the input intensity, the expected N-photon enhancements could potentially reach factors of over a million, given the achieved 1000-fold intensity enhancements by wavefront shaping[9]. A simultaneous correction of temporal distortions would result in an additional gain in signal and could pave the way to coherent control of nonlinear processes in scattering media[24, 25, 27, 28]. Beyond these practical aspects, the question to what extent can one control an optical pulse propagating through a random medium is a fundamental wave-propagation and control question, with fascinating analogues in acoustics and radio-frequency (RF) electromagnetic waves [29-31].

In this work we demonstrate the ability to control and correct, in both space and time, distorted ultrashort pulses after propagation through a scattering medium. We correct both spatial and temporal distortions simultaneously by manipulating only the spatial degrees of freedom of the incident wavefront. The key to this degree of control is the optimization of a nonlinear signal, two-photon fluorescence in our case. Naively, one would expect that in order to effectively cancel spatial and temporal distortions, it is required to have control over both the spatial *and temporal* degrees of freedom of the incoming wavefront, as is the case in time-reversal in acoustics and RF[29-31]. Indeed, temporal pulse shaping techniques[32] demonstrated recently the ability to compress the temporal duration of the light at a single speckle in the scattered field[19], without alleviating the spatial distortions. Here we show that, surprisingly, *spatial control alone* is sufficient for simultaneously correcting both the *spatial and temporal* distortions. The deep reason being that scattering couples the spatial and temporal degrees of freedom.

One of the striking results in the works of Vellekoop et al. is that the optimized focal spot can be smaller than the original focal spot without a scatterer[11, 31]. Here we show that our optimization scheme not only corrects for distortions induced by scattering, but it can lead to pulses which are shorter even than the input pulse (if that pulse was not Fourier limited). These results hold great potential for many nonlinear optics applications in scattering media, such as multiphoton microscopy[21], coherent quantum control[23], optical trapping[33], and nano-surgery. As a demonstration, we apply our technique to refocus 100fs transform-limited (TL) pulses through a multiply scattering 1mm thick rat brain tissue.

## Results

**Simultaneous focusing and pulse compression by wavefront shaping**

The experimental setup for spatiotemporal focusing is presented in Fig. 1(a). An ultrashort pulse was focused through a random scattering medium to a two-photon fluorescence (2PF) screen. The incoming wavefront phase was adaptively optimized using a two-dimensional SLM[34] to maximize the nonlinear 2PF at a selected point on the screen (see Methods). Imaging the 2PF after optimization revealed that the optimized 2PF was enhanced and refocused at the optimized spot (Fig.1 b-c). The refocused 2PF was not only localized in the transverse dimensions but was also confined along the axial dimension, as verified by imaging different depths in the 1mm thick 2PF screen (Fig.1 d-e). The axial confinement of the 2PF is obtained in the same manner that optical sectioning is achieved in 2PF microscopy[21]. Most importantly, as we show below, this spatial focusing is accompanied by significant temporal focusing as well, even though no special attempt has been made to control the temporal degrees of freedom.

To investigate the temporal properties of the scattered fields and to prove temporal compression, we have devised the characterization setup presented in Fig. 2(a). The key element in this setup is a Michelson interferometer added before the SLM, which produces spatially-resolved 2PF autocorrelation[35]. With this setup we could then characterize the temporal profile of the light fields *at each and every point in the image*. By scanning the pulse separation $\tau$ while collecting 2PF images with the EMCCD, we extract the pulse autocorrelation at each point in the image simultaneously. To demonstrate that our optimization scheme does not only correct for the temporal distortions induced by scattering but can lead to pulses shorter than the input pulse, we chirped the transform-limited 100fs laser pulses to ~400fs by passing them through a 152mm-long slab of F3 glass. Figure 2(b) shows that before any optimization, the autocorrelation of the light emerging from the scattering medium is ~850 fs long at all points (1/e width). We then applied the optimization algorithm (blocking one of the arms of the interferometer or setting its delay to zero) and then repeated the temporal measurement with the optimized field. Our results show that by simply optimizing a 2PF using only the spatial degrees of control we obtained a refocused pulse in both space (Fig. 1b-e) and time (Fig. 2c-d). Note that the ~300fs refocused pulse autocorrelation at the optimization point is not only significantly shorter than the non-optimized scattered pulse, but is *shorter than the original chirped input pulse* (710fs 1/e width, Fig. 2d), proving pulse compression by wavefront controlled random scattering. The pulse at the optimization point is of shorter duration than any other point in the image (Fig. 2c). The data collected by our experimental system allows the complete spatiotemporal reconstruction of the scattered and optimized light fields (Fig. 2e-f, see Supplementary video 1). We note that the setup presented in Fig. 2a is required only for characterization purposes, and the simple setup of Fig.1a is sufficient for achieving the spatiotemporal focus. The measured optimized 2PF intensity was enhanced 15-fold compared to the average 2PF before optimization (Fig.1b-d). However, the actual enhancement of the axially confined 2PF is much larger since for the non-optimized case off plane fluorescence contributions from the 1mm thick 2PF screen are significant. Taking into account the off-plane contributions, we estimate the localized 2PF enhancement to be close to 800. This is consistent with a measured 20-fold gain of the excitation field intensity, and further experiments with thin fluorescent screens (see Supplementary Figure 3).

To understand how the scattered light field focuses in both space and time even though no effort is made to explicitly minimize the width of the focus or the pulse temporal width, one needs to consider the dependence of the optimized 2PF signal on the spatial and temporal distributions of the light field. Assuming a two-photon absorption spectrum which is wider than the pulse bandwidth, the 2PF intensity at any point (x,y,z) in space is given by: $I_{TPF}(x,y,z) \propto \int I^2(x,y,z,t)dt$, where $I(x,y,z,t)$ is the excitation field intensity at the point (x,y,z) at time t. Thus, as the input pulse energy $E = \int I(x,y,z,t)dtdxdydz$ is given, the maximum 2PF is obtained for the field which is focused to the smallest possible spatial extent and the shortest temporal duration. This approach is in the spirit of previous works optimizing a nonlinear signal for temporal compression alone[36], and the recent works optimizing a linear signal for perfect spatial focusing of a monochromatic source[11]. Our work combines these two independent approaches to a direct model-free method for spatiotemporal focusing in random media.

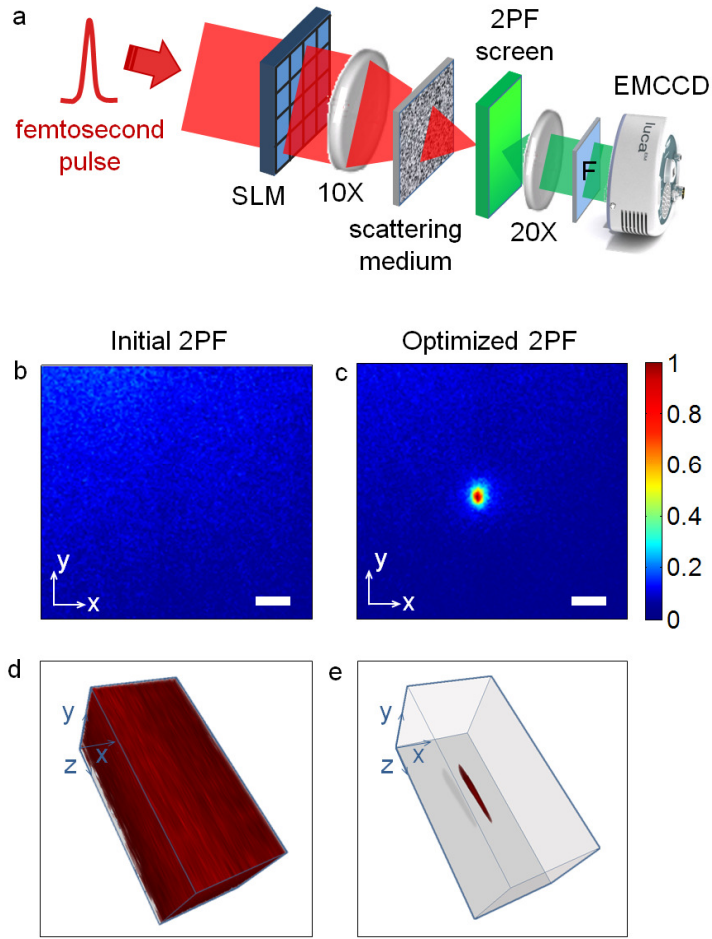

Figure 1: Spatiotemporal focusing by optimizing a two-photon fluorescence (2PF) signal: (a) the experimental setup: an ultrashort pulse is focused to a 2PF screen placed behind a scattering medium. A two-dimensional SLM controls the incident wavefront, optimizing the 2PF at a selected point in the screen, imaged by an EMCCD, (F - band-pass filter). (b-c) 2PF images before (b) and after (c) optimization at the optimized plane (x-y), demonstrating spatial refocusing and a 15-fold gain in the *measured* 2PF. (c-d) Depth resolved images of the 2PF before (d) and after (e) optimization, showing that the optimized TPF is confined along the axial (z) dimension, in addition to the transverse confinement. Taking into account the axial confinement, the localized 2PF enhancement is estimated to be approximately 800 (see Supplementary Figures 2,3). The pulse at the optimization point is also temporally compressed as is shown in Fig. 2. scale-bars in (b,c) are 25μm; rendered x-y-z field in (d,e) is 190x190x400μm.

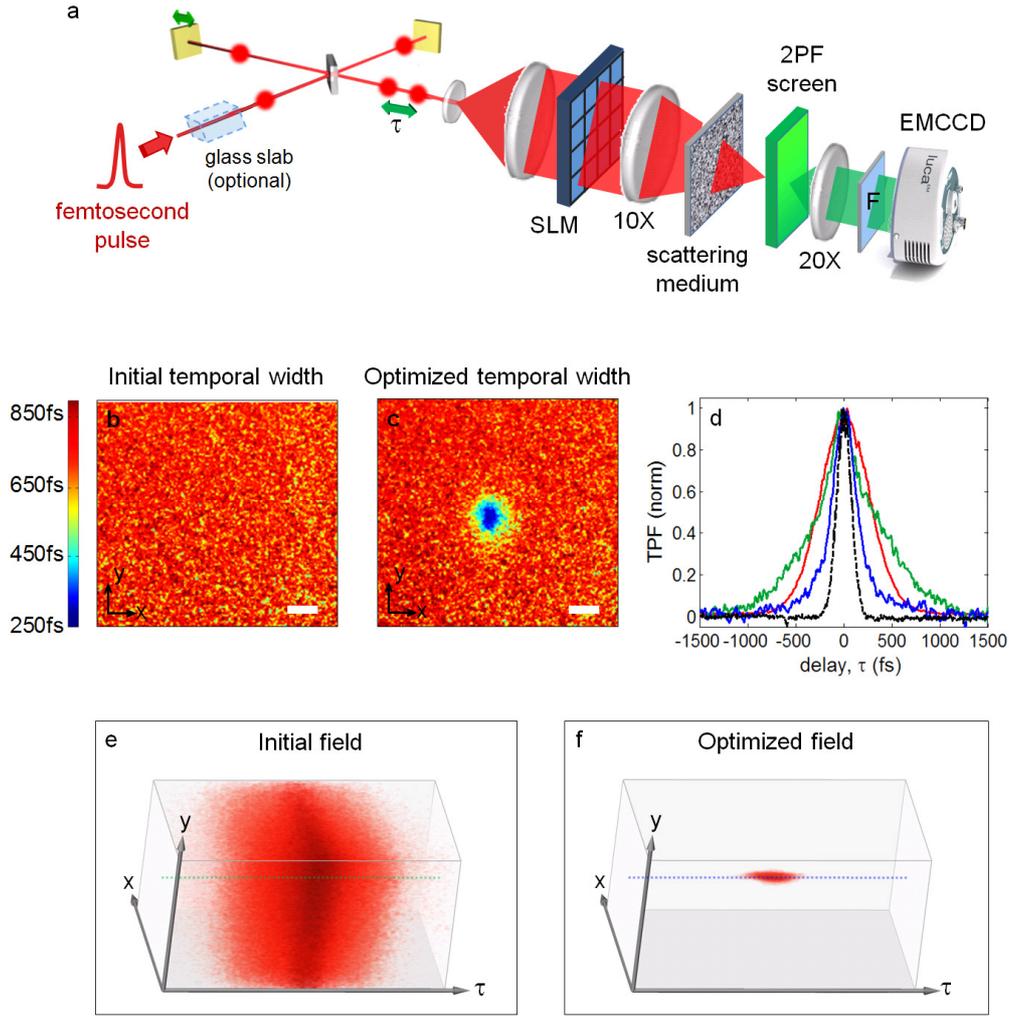

Figure 2: Spatiotemporal characterization of the scattered fields: (a) experimental setup: utilizing a Michelson interferometer while imaging the two-photon fluorescence (2PF), spatially resolved autocorrelation is measured on the entire field simultaneously. A 152mm long glass slab was used to pre-chirp the input pulse. (b-c) Maps of the temporal 1/e width of the spatially resolved autocorrelation, showing temporal compression of the optimized pulse at the optimization point. (d) Measured fringe-averaged autocorrelation of the chirped input pulse (red, 1/e width=710fs), the non-optimized scattered pulse (green, 1/e width=900fs), and the optimized pulse (blue, 1/e width=365fs), demonstrating that the optimized pulse is *shorter than the input pulse,* using only spatial wavefront shaping. The transform-limited pulse autocorrelation is plotted in dashed black for comparison (1/e width=210fs); (e,f): Rendering of the spatially-resolved autocorrelations for the non-optimized and optimized fields, revealing a dispersed spatiotemporal speckle in the non-optimized case (e), and a localized 2PF in both space and time in the optimized case (f) (see Supplementary Video 1). Dashed green/blue lines in (e-f) are the locations of autocorrelations plotted in corresponding colors in (d); rendered spatiotemporal volume is 200μm x 200μm x 2200fs ; scale-bars in (b,c) are 25μm;

The fact that the 2PF signal is optimized by the shortest pulse does not necessarily explain how such a pulse can be obtained by controlling solely the wavefront spatial phase. To emphasis this we note that every single pixel in our SLM can only induce minute delays of up to 3fs, whereas the pulse autocorrelation was shown to be compressed from a full width at half max of 590fs to 270fs (Fig. 2d). To achieve such temporal control one has to gain control over the different spectral phases in the pulse bandwidth as is done in conventional Fourier-domain pulse shapers (Fig. 3a)[32]. The spectral phase control must have a spectral resolution, $\Delta f$, smaller than $1/\tau_{max}$, where $\tau_{max}$ is the maximum controllable delay. This is usually achieved in a 4-f pulse-shaper by scattering from a diffraction grating (Fig. 3a), such that every SLM pixel controls the phase of a narrow spectral band with bandwidth $\Delta f < 1/\tau_{max}$. The spectral resolution, $\Delta f$, is determined by the grating period and geometry of the system[32]. Wavefront shaping in a random scattering medium is analogous to a Fourier domain pulse shaper in the sense that in both cases scattering couples the spatial and spectral/temporal degrees of freedom. Instead of using scattering from an ordered grating, the scattering in a disordered sample couples each SLM pixel to a temporally speckled field after the sample. These fields form an effective basis that can span a desired temporal function, in the same manner the spectral basis is used in a pulse-shaper (Fig. 3b). By controlling the relative phases of these random trains of pulses, they can be made to add up coherently at any desired instance. Thus, the controlled spectral resolution is $\Delta f=1/\tau_{max}$ where $\tau_{max}=\Delta L/c$ is the maximum relative temporal delay in the sample, given by the optical paths length differences $\Delta L$ divided by the speed of light, c.

A scattering medium in conjunction with an SLM can thus be thought of as a temporal pulse shaper with an extremely high temporal dynamic range. This is the mechanism which allowed the temporal compression of a chirped input pulse using only the spatial degrees of freedom (Fig. 2d) [37]. Strikingly, the more temporally complex the random scattering is the better is the temporal control attainable, given enough degrees of spatial control[31] (see Supplementary Figure 4). In our optimization process the phases of the temporally speckled fields are adjusted so they coherently add at one specific, though not predetermined instance, forming a short pulse. An approach for optimizing a single specific predetermined point in time has been recently suggested by Aulbach et al.[18].

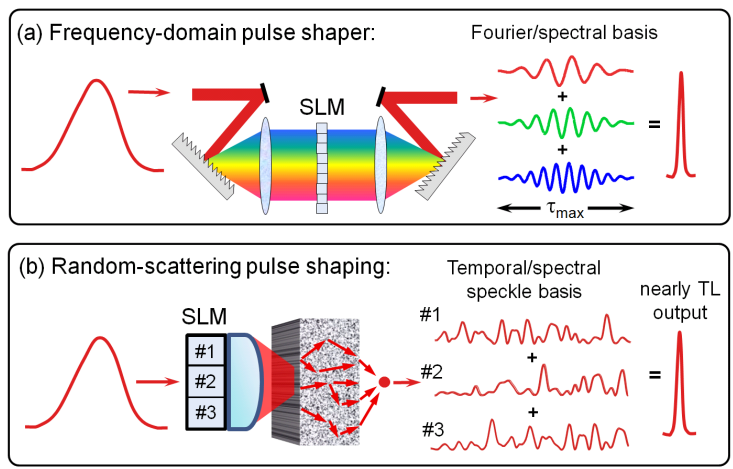

Figure 3: Mechanism for temporal compression using only spatial degrees of freedom and random scattering: much like a Fourier pulse shaper[32] (a), where the spatial SLM's pixels are coupled to the spectral degrees of freedom by scattering from a grating; coherent scatterings in a random medium (b) couple each SLM pixel to a different linear combination of the spectral (also temporal) degrees of freedom. Forming a new random spectral basis that is phase-controlled by the SLM. In both cases, the maximum controllable delay $\tau_{max}$, dictates the shaping spectral resolution $\Delta f=1/\tau_{max}$, and is determined by the optical paths lengths differences in the medium/shaper.

**Correction of spatiotemporal distortions in practical applications**

Our results suggest that spatial wavefront control is sufficient for correcting both the spatial and temporal distortions of ultrashort pulses in scattering media. This conclusion holds great promise for applications employing femtosecond pulses such as multiphoton microscopy, laser writing, coherent control and optical manipulation. What spatiotemporal distortions should one expect in a given application? Clearly, considerable distortions of the pulse temporal profile would only occur if the scattering induces optical path length differences $\Delta L$ that are larger than the pulse width: $\Delta L > c \cdot \tau_{TL}$. However, complex spatial distortions can be present even without significant temporal distortions[38]. The exact regime of spatiotemporal distortions in a specific scenario needs to be carefully evaluated by considering the sample parameters, i.e. thickness and scattering length[3, 4], as well as the geometrical parameters of the system, which can yield temporal distortions even from a thin scattering surface[2, 38, 39]. Our experimental system is unique in the sense that it allows for direct determination of the spatio-temporal distortions and their correction.

As a demonstration, we study the spatiotemporal distortions in a challenging multiple-scattering scenario of focusing 100fs pulses through a 1mm thick brain tissue, a task which is relevant to biophotonic applications. The results of this study are presented in Fig. 4. Imaging the 2PF from a 100fs TL pulse focused through a 1mm thick rat brain tissue reveals that this sample significantly distorts the spatial profile of the focused beam (Fig. 4a), as expected from a thick scattering medium. However, by measuring the spatially resolved 2PF autocorrelation, we find that the sample does not appreciably temporally distort the 100fs TL pulses over the entire speckled pattern in the 120x120µm imaged field (Fig.4c). This result is consistent with the scattering and geometrical parameters of this focusing scenario; Although the typical scattering lengths in brain tissues are 100-200µm in the near-IR[8, 22, 40], because of the low index contrast of biological tissues scattering is predominantly in the forward direction[22]. From the maximum spatial spread of the speckled 2PF pattern at the focal plane (~±50µm, Fig. 4a) one can deduce that most of the geometrical path length differences to the focus are smaller than the 100fs pulse width, resulting in only minor temporal distortions. Of course, the results may differ when shorter pulses are employed. Regardless of the spatiotemporal distortions, applying our adaptive optimization results in correction of all distortions in the focal spot, and in refocusing of the pulse in space and time (Fig.4 b,c). After optimization, the measured 2PF is increased 30-fold compared to the brightest 2PF before optimization, and the pulse autocorrelation at the focus is identical to the TL pulse autocorrelation (Fig. 4c).

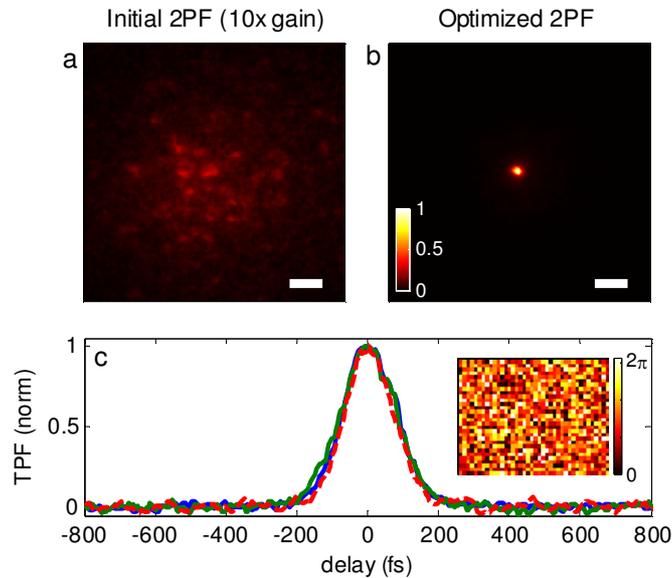

Figure 4: Spatiotemporal focusing of 100fs TL pulses through 1mm thick brain tissue: two-photon fluorescence (2PF) before (a) and after (b) optimization, showing a 30-fold increase in 2PF relative to the maximum 2PF intensity before optimization. The 2PF image before optimization is shown with 10x display gain; Scale-bars, 15µm. (c) Spatially resolved autocorrelation of the initial non-optimized pulse at the center (blue) and top-right corner of the field (green), both with FWHM=180±10fs, and autocorrelation of the optimized pulse (dashed red, FWHM=170fs). The optimized pulse autocorrelation is identical to the transform limited pulse shown in Fig.2d. Inset: SLM phase pattern used to generate the corrected optimized spot.

**Discussion**

We have presented a direct approach for characterizing and correcting spatio-temporal distortions in random scattering media. By optimizing a two-photon fluorescence signal an arbitrary pulse can be compressed inside a scattering sample[10]. Tailoring a complex femtosecond function would be possible by adaptive optimization of other nonlinear processes such as coherent anti-Stokes Raman scattering[26] or photo-reaction products[24, 25]. Furthermore, recent results suggest that the scattering-aided spatial confinement of the shaped pulse may be better than that possible by the optical system without the scattering medium[11], potentially achieving sub-wavelength focusing[27, 28].

Our adaptive technique is robust and is able to follow slow dynamics of non static samples (see Methods). However, its current implementation is too slow to be used in real-time or in-vivo imaging applications. The main limitations in our setup are the SLM refresh rate and the slow EMCCD image acquisition, both of which may be improved by using faster devices; it is also likely that improvements can be achieved using faster optimization algorithms. Combining our technique with temporal focusing[39] may allow deep nonlinear imaging in scattering tissue, enhancing the signal while rejecting out-of-focus contributions[40]. Applications in imaging are especially attractive when the same correction is usable over the entire field of interest, allowing scanning of the optimized spot[12]. This is the case when the inhomogeneities are substantially above the focal plane, e.g. the skin and cartilage overlying the brain[8]. An intriguing question to be investigated concerns what information on the scattering medium may be extracted from the optimal phase function.

# References

1. Sebbah, P. Waves and imaging through complex media. (Kluwer Academic Publishers, Dordrecht ; Boston; 2001).
2. Tal, E. & Silberberg, Y. Transformation from an ultrashort pulse to a spatiotemporal speckle by a thin scattering surface. *Opt Lett* **31**, 3529-3531 (2006).
3. Bruce, N.C. et al. Investigation of the temporal spread of an ultrashort light pulse on transmission through a highly scattering medium. *Appl Opt* **34**, 5823-5828 (1995).
4. Szmacinski, H., Gryczynski, I. & Lakowicz, J.R. Spatially localized ballistic two-photon excitation in scattering media. *Biospectroscopy* **4**, 303-310 (1998).
5. Dela Cruz, J.M., Pastirk, I., Comstock, M., Lozovoy, V.V. & Dantus, M. Use of coherent control methods through scattering biological tissue to achieve functional imaging. *PNAS* **101**, 16996-17001 (2004).
6. Tyson, R.K. Principles of adaptive optics, Edn. 2nd. (Academic Press, Boston; 1998).
7. Nature Photonics Technology Focus: Adaptive optics. *Nat Photon* **5**, 15-28 (2011).
8. Rueckel, M., Mack-Bucher, J.A. & Denk, W. Adaptive wavefront correction in two-photon microscopy using coherence-gated wavefront sensing. *PNAS* **103**, 17137-17142 (2006).
9. Vellekoop, I.M. & Mosk, A.P. Focusing coherent light through opaque strongly scattering media. *Opt Lett* **32**, 2309-2311 (2007).
10. Vellekoop, I.M., van Putten, E.G., Lagendijk, A. & Mosk, A.P. Demixing light paths inside disordered metamaterials. *Opt Express* **16**, 67-80 (2008).
11. Vellekoop, I.M., LagendijkA & Mosk, A.P. Exploiting disorder for perfect focusing. *Nat Photon* **4**, 320-322 (2010).
12. Vellekoop, I.M. & Aegerter, C.M. Scattered light fluorescence microscopy: imaging through turbid layers. *Opt Lett* **35**, 1245-1247 (2010).
13. Popoff, S.M. et al. Measuring the transmission matrix in optics: an approach to the study and control of light propagation in disordered media. *Phys Rev Lett* **104**, 100601 (2010).
14. Popoff, S., Lerosey, G., Fink, M., Boccara, A.C. & Gigan, S. Image transmission through an opaque material. *Nat Commun* **1**, doi:10 1038/ncomms1078 (2010).
15. Cizmar, T., Mazilu, M. & Dholakia, K. In situ wavefront correction and its application to micromanipulation. *Nat Photon* **4**, 388-394 (2010).
16. Yaqoob, Z., Psaltis, D., Feld, M.S. & Yang, C. Optical phase conjugation for turbidity suppression in biological samples. *Nat Photon* **2**, 110-115 (2008).
17. Cui, M. & Yang, C. Implementation of a digital optical phase conjugation system and its application to study the robustness of turbidity suppression by phase conjugation. *Opt. Express* **18**, 3444-3455 (2010).
18. Aulbach, J., Gjonaj, B., Johnson, P.M., Mosk, A.P. & Lagendijk, A. Control of light transmission through opaque scattering media in space and time. *arXiv:1011.5959v1* (2010).
19. McCabe, D.J. et al. Shaping speckles: spatio-temporal focussing of an ultrafast pulse through a multiply scattering medium. *arXiv:1101.0976* (2011).
20. Goodman, J.W. Speckle phenomena in optics : theory and applications. (Roberts & Co., Englewood, Colo.; 2007).
21. Denk, W., Strickler, J.H. & Webb, W.W. Two-photon laser scanning fluorescence microscopy. *Science* **248**, 73-76 (1990).
22. Oheim, M., Beaurepaire, E., Chaigneau, E., Mertz, J. & Charpak, S. Two-photon microscopy in brain tissue: parameters influencing the imaging depth. *J Neurosci Methods* **111**, 29-37 (2001).
23. Rabitz, H., de Vivie-Riedle, R., Motzkus, M. & Kompa, K. Whither the Future of Controlling Quantum Phenomena? *Science* **288**, 824-828 (2000).
24. Assion, A. et al. Control of chemical reactions by feedback-optimized phase-shaped femtosecond laser pulses *Science* **30**, 919-922 (1998).
25. Pearson, B.J., White, J.L., Weinacht, T.C. & Bucksbaum, P.H. Coherent control using adaptive learning algorithms. *Physical Review A* **63**, 063412 (2001).
26. Oron, D., Dudovich, N. & Silberberg, Y. All-optical processing in coherent nonlinear spectroscopy. *Physical Review A* **70**, 023415 (2004).
27. Stockman, M.I., Faleev, S.V. & Bergman, D.J. Coherent control of femtosecond energy localization in nanosystems. *Phys Rev Lett* **88**, 067402 (2002).
28. Aeschlimann, M. et al. Adaptive subwavelength control of nano-optical fields. *Nature* **446**, 301-304 (2007).
29. Fink, M. Time Reversed Acoustics. *Physics Today* **50**, 34-40 (1997).
30. Lerosey, G. et al. Time reversal of electromagnetic waves. *Phys Rev Lett* **92**, 193904 (2004).
31. Derode, A. et al. Taking advantage of multiple scattering to communicate with time-reversal antennas. *Phys Rev Lett* **90**, 014301 (2003).
32. Weiner, A.M. Femtosecond pulse shaping using spatial light modulators. *Rev. Sci. Instrum.* **71**, 1929 (2000).
33. Jiang, Y., Narushima, T. & Okamoto, H. Nonlinear optical effects in trapping nanoparticles with femtosecond pulses. *Nat Phys* **6**, 1005-1009 (2010).
34. Vellekoop, I.M. & Mosk, A.P. Phase control algorithms for focusing light through turbid media. *Optics Communications* **281**, 3071-3080 (2008).
35. Diels, J.C., Fontaine, J.J., McMichael, I.C. & Simoni, F. Control and measurement of ultrashort pulse shapes (in amplitude and phase) with femtosecond accuracy. *Appl Opt* **24**, 1270 (1985).
36. Yelin, D., Meshulach, D. & Silberberg, Y. Adaptive femtosecond pulse compression. *Opt Lett* **22**, 1793-1795 (1997).
37. Lemoult, F., Lerosey, G., de Rosny, J. & Fink, M. Manipulating spatiotemporal degrees of freedom of waves in random media. *Phys Rev Lett* **103**, 173902 (2009).
38. Small, E., Katz, O., Eshel, Y., Silberberg, Y. & Oron, D. Spatio-temporal X-wave. *Opt. Express* **17**, 18659-18668 (2009).
39. Oron, D., Tal, E. & Silberberg, Y. Scanningless depth-resolved microscopy. *Opt. Express* **13**, 1468-1476 (2005).
40. Wilt, B.A. et al. Advances in light microscopy for neuroscience. *Annu Rev Neurosci* **32**, 435-506 (2009).

## Acknowledgements

We thank Eduard Korkotian and Gayane Grigoryan for the brain sample preparation, Roee Ozeri for the EMCCD camera, Meital Covo for graphical design and Dan Oron for fruitful discussions. This work was supported by grants from the Israel Science Foundation, the Israel Ministry of Science, and the Crown Photonics Center.

## Author contribution

O.K., Y.B, E.S. and Y.S. designed the experiments. O.K., Y.B. and E.S. performed the experiments and analyzed the data. O.K., Y.B. and E.S. and Y.S. wrote the paper.

## Methods

### Experimental system

The complete experimental setup is shown in Fig.2a. The laser source is a Spectra-Physics Tsunami, producing 100fs pulses at 80MHz repetition rate with an average power of 2.1W. SLM is a Hamamatsu LCOS-SLM X10468-02. Microscope objectives used for focusing and imaging are 10X 0.25NA and 20X 0.56NA, respectively. 2PF screens are made from Disodium Fluorescein in ethanol. A 1mm thick cuvette was used as the 2PF screen in the experiments reported in Figs.1-2. A <100$\mu$m thick layer of Fluorescein placed between a microscope slide and a 100$\mu$m thick glass cover-slip was used in the experiment reported in Fig.4. In the imaging path, the 2PF was separated from the excitation field by a dichroic mirror (Semrock FF720-SDi01) and a band-pass filter (Chroma, D525-250). The 2PF was imaged by an EMCCD (Andor Luca-S) and the excitation field was simultaneously imaged by a CCD (Watec WAT-120N, not shown in Fig.2). The scattering sample used in the experiment reported in Figs.1-2 is a 60$\mu$m thick diffusive plastic tape placed 4mm from the 2PF cuvette. Optimization point in Figs.1-2 was 790$\mu$m from the optical axis. Autocorrelation traces were obtained by slowly scanning the delay line of the interferometer while rapidly dithering its position, utilizing a 1s-long integration time of the EMCCD to obtain a fringe-averaged autocorrelation trace[35].

### Optimization process

The SLM was divided to 1200 equally sized square segments, and the phase of the different segments was optimized using the iterative partitioning algorithm described in by Vellekoop et. al[34]. At each iteration the algorithm randomly selected half of the segments and adjusted their overall phase to maximize the target 2PF (see Supplementary Figure 1). The optimization continues indefinitely and follows the dynamics of the system. 1100 iterations were used in the experiment presented in Figs. 1-2 and 300 iterations were used in the experiment presented in Fig.4. The EMCCD integration time was 2s in the optimization presented in Fig.1-2 and 100ms-500ms in the optimization presented Fig.4 (integration time was adaptively lowered with signal enhancement). Before optimizing the 2PF in the experiment presented in Figs. 1-2, we ran 100 iterations with <50ms integration time optimizing the excitation field intensity.

### Brain sample preparation

A two week-old rat was decapitated and the brain was separated from the scalp. The separate brain was placed on a chopper device and the hippocampal region was cut in order to obtain several thin living brain sections of 500$\mu$m to 1000$\mu$m thickness. The sections were fixed in 4% paraformaldehyde during 1h, extensively washed with phosphate buffer and stored at +4 C. The samples were then placed on a glass microscope slide in a solution of 25% phosphate buffer and 75% Glycerol and sealed with a 100$\mu$m thick glass cover-slip.

# Supplementary Material

## 1. Optimization algorithm

The iterative algorithm used for optimizing the 2PF is the partitioning algorithm described by Vellekoop et al. [Optics Communications 281, 3071-3080 (2008)]. At each iteration half of the SLM segments were randomly selected. The phases of the selected segments were ramped from 0 to $2\pi$ in 21 equally spaced steps, and for each step the 2PF signal was recorded. As the pulse's bandwidth is much narrower than its central wavelength, the 2PF signal at the $r^{th}$ step closely follows: $I_r^{2PF} \propto \left[1 + \cos\left(\frac{2\pi r}{21} - \phi_{opt}\right)\right]^2$, where $\phi_{opt}$ is the optimal phase maximizing the 2PF signal. This is indeed the case as is shown in Supplementary Figure 1, which displays the measured 2PF as a function of the added phase for the 21 steps of the last iteration in the experiment reported in Figs.1-2. For each iteration we extracted the optimal phase $\phi_{opt}$, from the measured values $I_r^{2PF}$, by calculating the phase of the oscillating cosine with a $2\pi$ period, using to the following formula:

$$\phi_{opt} = \tan^{-1}\left(\frac{\sum_{r=1}^{21} I_r^{2PF} \sin\left(\frac{2\pi r}{21}\right)}{\sum_{r=1}^{21} I_r^{2PF} \cos\left(\frac{2\pi r}{21}\right)}\right)$$

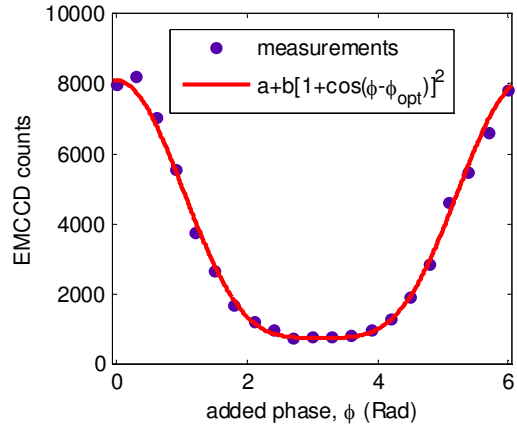

Supplementary Figure 1: Example for the measured 2PF signal at one iteration as a function of the added phase $\phi_r=2\pi r/21$, to the randomly selected SLM segments. The red line is a fit to a squared cosine function

## 2. Axial confinement and enhancement factor of the two-photon fluorescence signal

In all of our experiments the measured 2PF is the incoherent sum of the fluorescence contributions from all the different depths in the fluorescent screen. For a perfectly focused beam, only the volume confined by the confocal parameter of the beam contributes to the 2PF. This provides the axial sectioning of two-photon microscopy and is the case for the optimized 2PF, in contrast to the non-optimized 2PF which has no axial confinement. This is shown in Supplementary Figure 2, plotting the 2PF imaged at the different depths along the z-axis at the transverse (x-y) location of the optimized spot. This trace is a one-dimensional cross-section of the 3D plots of Fig.1 (d-e). It can be thus concluded that the measured 2PF in the optimized case originates from a much smaller volume than the 2PF in the non-optimized. The latter is the incoherent sum of significant out of focus contributions from the thick (1mm) 2PF screen.

Therefore, the localized enhancement of 2PF signal at the optimization point in the experiments presented in Figs. 1-2 cannot be obtained directly from the EMCCD image. However, the localized enhancement *of the excitation field* can be measured directly by *coherently* imaging the optimization point on an additional CCD camera. We have carried out this measurement which yielded an enhancement of 20 in the excitation field intensity. This represents a 2PF enhancement of no less than $20^2=400$ without taking into account temporal compression. Taking into account the ~2x temporal compression this result yields a ~800-fold enhancement in the localized 2PF. This enhancement factor agrees with the 15-fold 2PF enhancement measured directly by the EMCCD camera times the ratio of the 2PF screen thickness to the Rayleigh range of the optimized ~7μm diameter spot.

To further verify that the local 2PF signal is enhanced by three orders of magnitude, we carried out an optimization experiment using a thin (<100μm) 2PF screen and a TL input pulse. Supplementary figure 3 shows the values of 2PF intensity versus the excitation intensity for each iteration. As expected, plotting the 2PF intensity enhancement as a function of the optimized excitation intensity reveals a nearly quadratic relation, and a *measured* 2PF enhancement of ~1000 compared to the average 2PF before optimization (Supplementary Figure 3a). Repeating the same experiment with a 1mm thick 2PF screen yields a much lower measured enhancement in the 2PF as result of the off-plane contributions, masking the optimized 2PF for low enhancement values (Supplementary Figure 3b). This is confirmed by the nearly quadratic relation between the 2PF and the excitation intensity which is obtained just for high enhancement values (Supplementary Figure 3b).

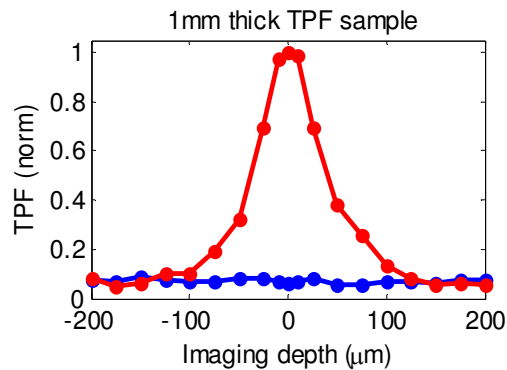

Supplementary Figure 2:.depth cross-section of the two-photon fluorescence in the optimized (red) and non-optimized (blue) cases, showing the axial confinement of the optimized 2PF and the off-axis background contributions in the non-optimized case.

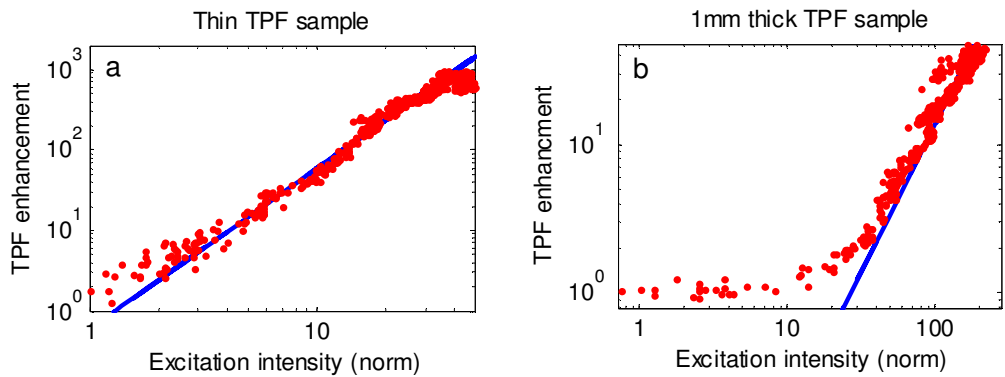

Supplementary Figure 3: Comparison of the dependence of the measured 2PF to the optimized excitation intensity during the optimization using a thin 2PF screen (a) and using a thick 2PF screen (b). Quadratic dependence is given by blue lines. In the thick screen case (b) the weak non-optimized signal is masked by significant off-plane fluorescence contributions, which masks the true enhancement of the localized 2PF.

## 3. Pulse compression as a function of the number of degrees of control

To study the pulse compression as a function of the number of degrees of spatial control (SLM segments), we have written a numerical simulation for the pulse evolution using the Fresnel-Huygens propagator under the paraxial approximation. The propagated field was calculated separately for each wavelength in the pulse bandwidth. The fields at the different frequencies were then summed to obtain the temporal profile at each point in space. We propagated the fields individually from each SLM segment to the optimized point of interest to create a broadband transfer matrix (impulse responses) for the different SLM pixels to the optimized point. The optimized SLM pattern for this matrix using the experimental parameters was calculated with the same algorithm used in the experiment, and then propagated through the system to generate the optimized pulse. The pulse autocorrelation width was then calculated for several simulation runs using different number of controllable SLM segments (the SLM was divided to equally sized controllable segments). The results show that increasing the number of degrees of control increases the attainable pulse compression (Supplementary Figure 4).

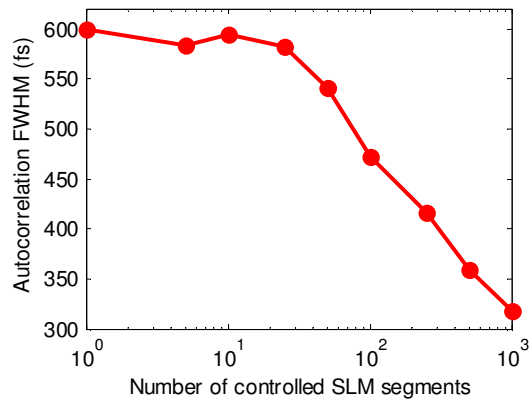

Supplementary Figure 4: Pulse compression as a function of the number of controlled degrees of freedom (SLM segments), showing increased pulse compression for more degrees of control (numerical simulation results).